\documentclass{aa}  
\usepackage{graphicx}
\usepackage{txfonts}
\usepackage{url}
\usepackage{multirow}
\usepackage{natbib}
\bibpunct[, ]{(}{)}{;}{a}{}{,}

\def\Lya{\mbox{Ly$\hspace{0.2ex}\alpha$}}
\def\Ha{\mbox{H$\hspace{0.1ex}\alpha$}}
\begin{document}

   \title{Dynamic \Lya\ jets}

   \author{J. Koza\inst{1}
          \and
          R. J. Rutten\inst{2,3}
	  \and
	  A. Vourlidas\inst{4}
          }

   \institute{Astronomical Institute, Slovak Academy of Sciences, 
              059 60 Tatransk\'a Lomnica, Slovakia
             \and
             Sterrekundig Instituut, Universiteit Utrecht, P.O. Box 80\,000,
             NL--3508 TA Utrecht, The Netherlands
             \and
             Institute of Theoretical Astrophysics, University of Oslo,
             P.O. Box 1029, Blindern, N--0315 Oslo, Norway
             \and
	     Code 7663, Naval Research Laboratory, Washington, DC 20375, USA
             }

   \date{Received DD MMMMMMM, 2008; accepted DD MMMMMMM, 2009}

%%%%%%%%%%%%%%%%%%%%%%%%%%%%%%%%%%%%%%%%%%%%%%%%%%%%%%%%%%%%%%%%%%%%%%%%
 \abstract
% context heading (optional), leave it empty if necessary  
   {The solar chromosphere and transition region are highly structured
   and complex regimes. A recent breakthrough has been the
   identification of dynamic fibrils observed in \Ha\ as caused by
   field-aligned magnetoacoustic shocks.}
   {We seek to find whether such dynamic fibrils are also observed in
   \Lya.}
   {We used a brief sequence of four high-resolution \Lya\ images of
   the solar limb taken by the Very high Angular resolution
   ULtraviolet Telescope (VAULT), which displays many extending and
%RE NO hyphen: proper name; NO comma, not extra info
   retracting \Lya\ jets. We measured their top trajectories and
   fitted parabolas to the 30 best-defined ones.}
   {Most jet tops move supersonically. Half of them decelerate,
   sometimes superballistically, the others accelerate. This
   bifurcation may arise from incomplete sampling of recurrent jets.}
  {The similarities between dynamic \Lya\ jets and \Ha\ fibrils
   suggest that the magnetoacoustic shocks causing dynamic \Ha\
   fibrils also affect dynamic \Lya\ jets.}
%%%%%%%%%%%%%%%%%%%%%%%%%%%%%%%%%%%%%%%%%%%%%%%%%%%%%%%%%%%%%%%%%%%%%%%%%


   \maketitle

%%%%%%%%%%%%%%%%%%%%%%%%%%%%%%%%%%%%%%%%%%%%%%%%%%%%%%%%%%%%%%%%%
\section{Introduction}
%%%%%%%%%%%%%%%%%%%%%%%%%%%%%%%%%%%%%%%%%%%%%%%%%%%%%%%%%%%%%%%%%

The solar chromosphere observed in \Ha\ is a bewildering mass of
elongated features, but a breakthrough concerning so-called dynamic
fibrils (henceforth \Ha\ DFs) has been made in the studies of
\citet{DePontieuetal2004,DePontieuetal2007a},
\citet{Hansteenetal2006}, \citet{RouppevanderVoortetal2007},
\citet{Hegglandetal2007}, and \citet{Langangenetal2008a,
Langangenetal2008b} following earlier work by
\citet{Suematsuetal1995}. These studies have established that \Ha\
DFs, which are rows of dark fibrilar features jutting out from plage
and network with periodic extension and retraction, display repetitive
mass loading by upward propagating magnetoacoustic shock waves driven
by the global solar oscillations. Reduction of the effective gravity
along tilted magnetic channels lowers their cutoff frequency and lets
them propagate into the chromosphere, steepen into shocks, and
repetitively lift the chromospheric-transition region interface.

These studies are all based on optical observations, but
\citet{deWijnandDePontieu2006} have studied transition-region jets in
\ion{C}{iv} in ultraviolet TRACE images and found remarkable
%RE I don't understand the comment
morphological similarities between \Ha\ DFs and their \ion{C}{iv}
counterparts. The DFs may represent the injection of cool material
postulated by \citet{Judge2008} as the source of hot sheaths making up
the transition region.

In this paper we report the presence of similar features in solar
\Lya\ images taken with the Very high Angular resolution ULtraviolet
Telescope \citep[VAULT,][]{Korendykeetal2001} during its second rocket
flight. VAULT acquired \Lya\ images with much higher spatial and
temporal resolution than the EUV imagery from the TRACE, SoHO, and
STEREO satellites but only for a few minutes.
%RJ %RE extra
The \Lya\ fine structure in these images is discussed by
\cite{Patsourakosetal2007} and \cite{JudgeandCenteno2008}. Here we
locate and study extending or retracting \Lya\ brightness structures
in images taken near the limb. We call these dynamic \Lya\ jets,
abbreviated to \Lya\ DJs, and compare them to \Ha\ DFs.

%RJ jet = long narrow feature, not necessarily dynamic
%RJ ``dynamic jet'' = those that extend or retract

%%%%%%%%%%%%%%%%%%%%%%%%%%%%%%%%%%%%%%%%%%%%%%%%%%%%%%%%%%%%%%%%%%%%%%%%
% Fig 1  image
%%%%%%%%%%%%%%%%%%%%%%%%%%%%%%%%%%%%%%%%%%%%%%%%%%%%%%%%%%%%%%%%%%%%%%%%
   \begin{figure} 
   \centering
   \resizebox{\hsize}{!}{\includegraphics{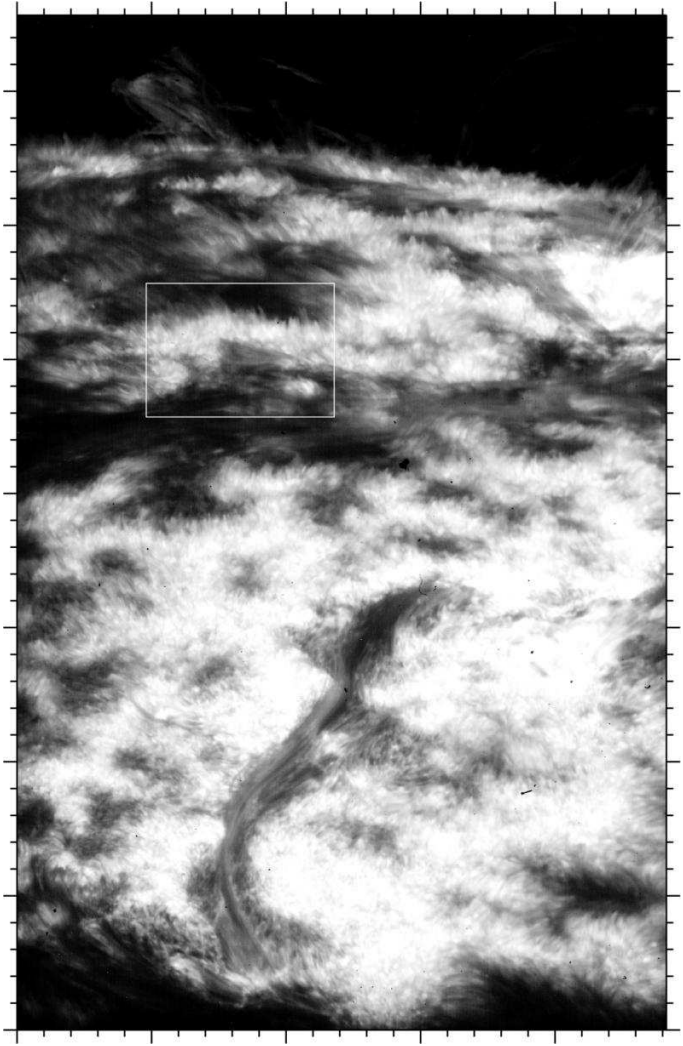}}
   \caption[]{
   \Lya\ image taken at 18:16:56\,UT covering the east limb. We invite
   the reader to inspect this image at large magnification with a pdf
   viewer. \Lya\ jets are grouped in hedgerows at network boundaries
   that are best seen in projection against the dark internetwork. The
   rectangle selects the hedgerow enlarged in Fig.~\ref{fig:closeup}.
   The contrast is enhanced by logarithmic brightness scaling.
   Tickmark spacing: 10\,arcsec.
}  \label{fig:image} 
\end{figure}
%%%%%%%%%%%%%%%%%%%%%%%%%%%%%%%%%%%%%%%%%%%%%%%%%%%%%%%%%%%%%%%%%%%%%%%%

%%%%%%%%%%%%%%%%%%%%%%%%%%%%%%%%%%%%%%%%%%%%%%%%%%%%%%%%%%%%%%%%%%%%%%%%
% Fig 2  closeup
%%%%%%%%%%%%%%%%%%%%%%%%%%%%%%%%%%%%%%%%%%%%%%%%%%%%%%%%%%%%%%%%%%%%%%%%
   \begin{figure}
   \centering
   \resizebox{\hsize}{!}{\includegraphics{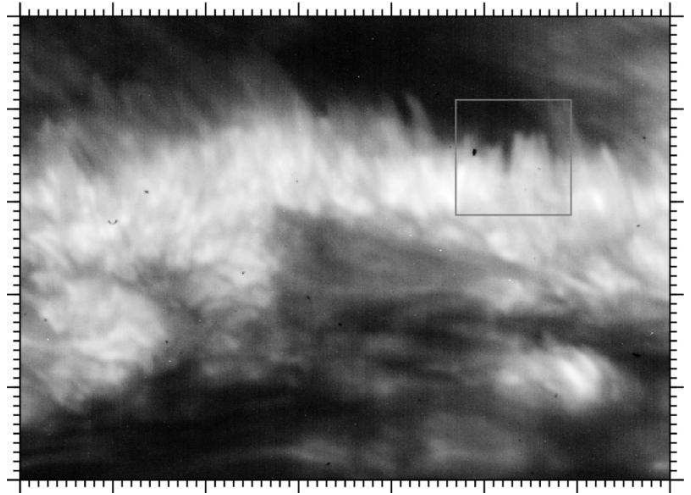}}
   \caption[]{
     Enlargement of the subfield outlined in Fig.~\ref{fig:image}. The
     rectangle outlines the smaller subfield shown in
     Fig.~\ref{fig:evolution} with \Lya\ DJs 3 and 11. Tickmark
     spacing: 1\,arcsec.
}  \label{fig:closeup}
   \end{figure}
%%%%%%%%%%%%%%%%%%%%%%%%%%%%%%%%%%%%%%%%%%%%%%%%%%%%%%%%%%%%%%%%%%%%%%%%

%%%%%%%%%%%%%%%%%%%%%%%%%%%%%%%%%%%%%%%%%%%%%%%%%%
% Fig 3 55DJs 
%%%%%%%%%%%%%%%%%%%%%%%%%%%%%%%%%%%%%%%%%%%%%%%%%%
   \begin{figure}
   \centering
   \resizebox{\hsize}{!}{\includegraphics{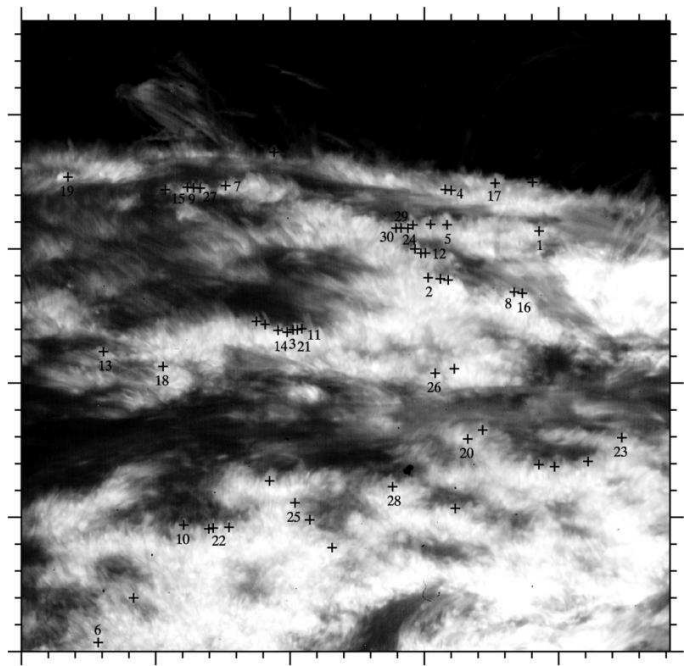}}
   \caption[]{
     The upper part of Fig.~\ref{fig:image} with 55 \Lya\ DJs marked
     by plus signs. The 30 DJs selected for further analysis are
     labeled with their reference number. Tickmark spacing:
     10\,arcsec.
}  \label{fig:55DJs}
   \end{figure}
%%%%%%%%%%%%%%%%%%%%%%%%%%%%%%%%%%%%%%%%%%%%%%%%%%%%%%%%%%%%%%%%%%%%%%%%

%%%%%%%%%%%%%%%%%%%%%%%%%%%%%%%%
% Fig 4 evolution
%%%%%%%%%%%%%%%%%%%%%%%%%%%%%%%%
   \begin{figure}
   \centering
   \resizebox{\hsize}{!}{\includegraphics{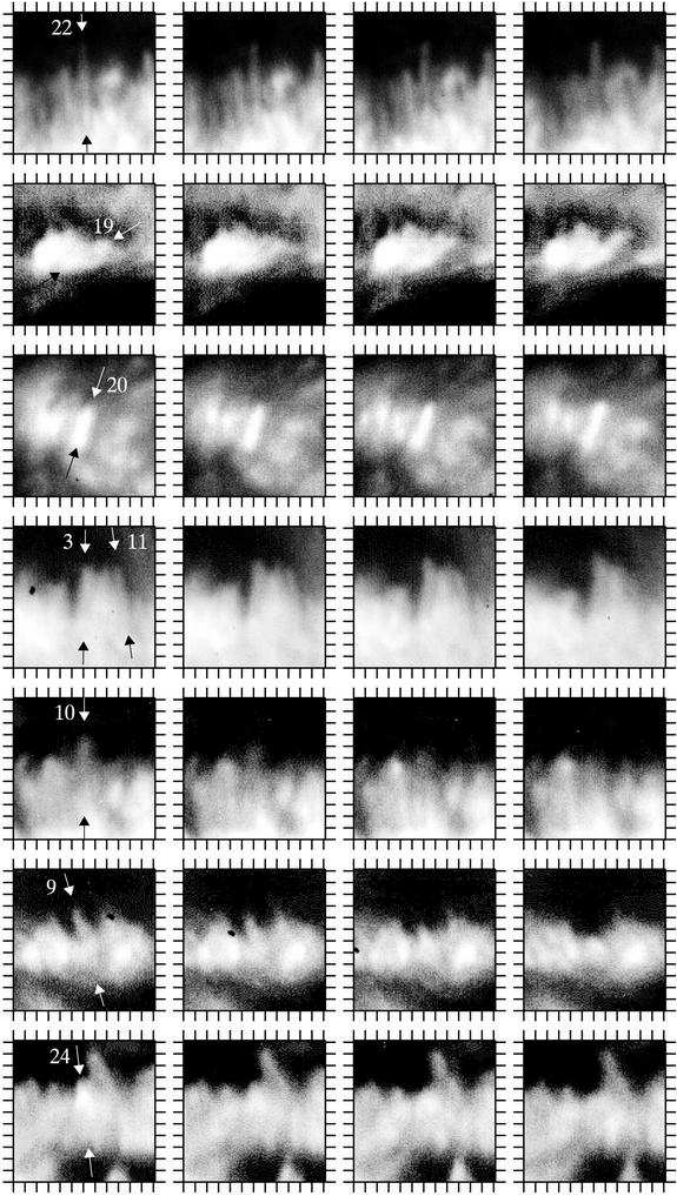}}
   \caption[]{
     Time evolution of extending (nrs.~22, 19, 20, 3) and retracting
     (nrs.~11, 10, 9, 24) \Lya\ DJs. The time increases with 17-s
     intervals between the columns. The arrows define the jet
     axis. The tickmarks have 1-arcsec spacing.
}   \label{fig:evolution}
   \end{figure}
%%%%%%%%%%%%%%%%%%%%%%%%%%%%%%%%%%%%%%%%%%%%%%%%%%%%%%%%%%%%%%%%%%%%%%%%

%%%%%%%%%%%%%%%%%%%%%%%%%%%
% Fig 5 intensity profiles
%%%%%%%%%%%%%%%%%%%%%%%%%%%
   \begin{figure}
   \centering
   \resizebox{0.967\hsize}{!}{\includegraphics{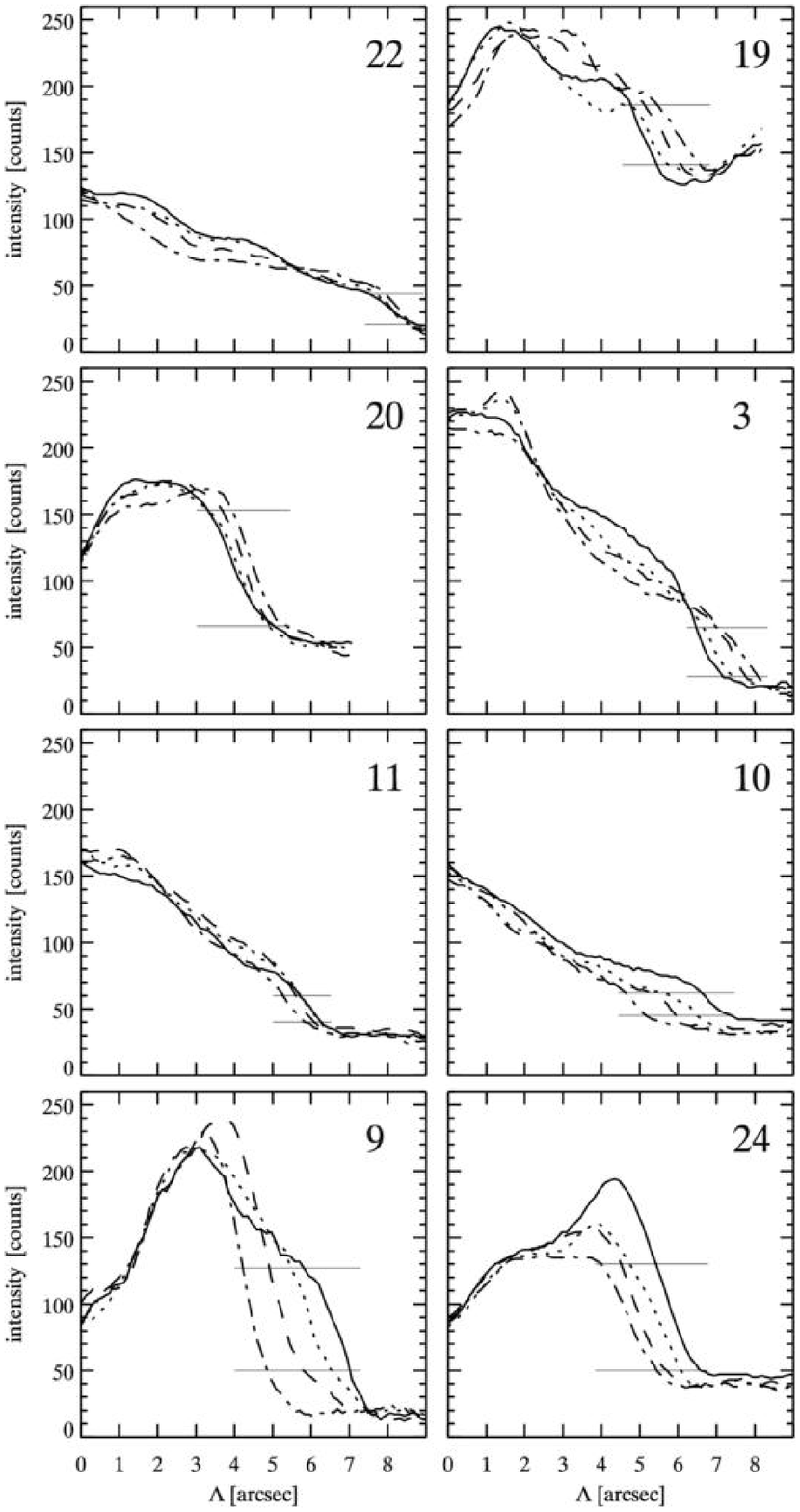}}
   \caption[]{
     Intensity profiles along the longitudinal axes of the \Lya\ DJs
     shown in Fig.~\ref{fig:evolution}. The solid, dotted, dashed, and
     dashed-dotted curves correspond to the 1st, 2nd, 3rd, and 4th
     panel in the corresponding row of Fig.~\ref{fig:evolution},
     respectively, and show the temporal evolution. The thin-line
     markers specify the ranges of intensity thresholds defining each
     jet top. The reference footpoints with $\Lambda=0$ are fixed in
     time.
}  \label{fig:profiles}
   \end{figure}
%%%%%%%%%%%%%%%%%%%%%%%%%%%%%%%%%%%%%%%%%%%%%%%%%%%%%%%%%%%%%%%%%%%%%%%%

%%%%%%%%%%%%%%%%%%%%%%%%%%%%%%%%%%%%%%%%%%%%%%%%%%%%%%%%%%%%%%%%%%%%%%%%
\section{Observations and measurements}
%%%%%%%%%%%%%%%%%%%%%%%%%%%%%%%%%%%%%%%%%%%%%%%%%%%%%%%%%%%%%%%%%%%%%%%%

We used \Lya\ 
%RJ %RE footnote extra; in an other A&A paper the language editor liked
%RJ such as footnote so the same here.
images\footnote{The VAULT data archives are available at
\url{http://wwwsolar.nrl.navy.mil/rockets/vault/}.}  recorded during
the second VAULT flight of 14 June 2002. The cadence was 17\,s, the
exposure time 1\,s, the image scale 0.124\,arcsec\,px$^{-1}$, and the
angular resolution 1/3~arcsec. The VAULT gratings isolated a 150\,\AA\
wide spectral band around \Lya\ and a narrow-band filter reduced the
passband to 72\,\AA\ FWHM. The resulting signal is about 95\% pure
\Lya\ emission (Teriaca \& Sch\"uhle, personal communication). More
detail is given in \citet{Korendykeetal2001} and
\citet{Patsourakosetal2007}.

The VAULT-II flight recorded two \Lya\ image sequences, a
seventeen-image one stepping over an extended active area on the disk
and a four-image one of quieter areas near the limb. We use only the
four limb images because \Lya\ jets are best observed as bright
features in projection against the dark internetwork background. The
four images were precisely co-aligned through cross-correlation using
routines of P.~S\"utterlin. We then applied appropriate greyscaling to
enhance their fine structure and used extensive visual inspection to
identify \Lya\ jets and to study their temporal behavior. They are
similar to limb spicules in being bright against dark internetwork and
in showing hedgerow-clustering at network boundaries. Similar \Lya\
jets are also seen in the disk sequence, but we decided to limit our
measurements to the limb sequence that shows them best. It clearly
shows temporal variation in the spatial extent of many \Lya\ jets,
even during the brief one-minute sequence duration.

The four images used here were taken between 18:16:56 and 18:17:47~UT
at the east limb. The first is shown in Fig.~\ref{fig:image}. Towards
the limb it displays thick hedgerows of bright \Lya\ jets, jutting out
at network boundaries. They are remarkably similar, at reversed
contrast, to the dark hedgerows in filtergrams taken in the wings of
\Ha\ (e.g., Fig.~9.1 of \citealp{2004suin.book.....S} and Fig.~7 of
\citealp{Rutten2007}), in which Dopplershift of the line core into the
wings selects dynamic fibrils over the quiescent network-spanning
fibrils that dominate the scene at \Ha\ line center. Such extended
fibrils appear dark in the VAULT-II images and provide the background
against which the hedgerows appear bright. Figure~\ref{fig:closeup}
shows an enlargement of one such hedgerow. The individual jets have
widths of only one arcsec or less, requiring VAULT's high resolution
for identification and tracking.

We have manually identified 55 \Lya\ jets that exhibit measurable
extension and/or retraction during the four-frame sequence, all
located in network clusters with convenient dark backgrounds. More
weeding, described below, resulted in the selection of 30 of these
\Lya\ DJs for presentation. They are marked by their reference number
in Fig.~\ref{fig:55DJs}. Some examples are shown in
Fig.~\ref{fig:evolution} in the form of small cutout sequences from
the four frames. There are many other features with similar morphology
in Fig.~\ref{fig:55DJs}; undoubtedly, many more might be recognized as
\Lya\ DJs in longer time sequences. However, there are also many
jet-like features that do not change at all during the brief sequence
duration. For example, the conspicuous jet to the right of retracting
\Lya\ DJ 24 in the bottom panels of Fig.~\ref{fig:evolution} does not
show any change.

The cutouts in Fig.~\ref{fig:evolution} illustrate that these \Lya\
DJs appear rather fuzzy, making measurement of their extension or
retraction somewhat imprecise. At the suggestion of the referee to an
earlier version of this paper, we developed a measuring method that
includes error estimation. We first defined reference footpoints and
longitudinal axis orientations for each of the 55 \Lya\ DJs to measure
their lengths. Since many jets have no clear base due to crowding in
the network from which they originate, the footpoint designation is
rather arbitrary but this does not affect our length change
measurements since we maintained the same footpoint for the three
other frames after selecting it in the first frame. We similarly
maintained the axis direction. No correction for projection effects
was applied. The limbward viewing suggests that, on average, these
\Lya\ DJs are viewed more or less from aside.

We then plotted the intensities along each jet against a length
coordinate $\Lambda$ measured in arcsec along the axis from the
footpoint at $\Lambda\!=\!0$. Figure~\ref{fig:profiles} shows the
intensity profiles for the same \Lya\ DJs as displayed in
Fig.~\ref{fig:evolution}. Each profile is box-car smoothed over
6~pixels (0.74\,arcsec) to reduce noise. The differences between the
four curves in each panel show the temporal jet evolution. The DJs in
the upper four panels extend; i.e., their outer parts become brighter.
Their lower parts sometimes weaken simultaneously (as for DJ~22 in the
first panel). The DJs in the lower four panels retract, without much
change in their lower parts. Inspection of all 55 plots made us
discard 15 DJs for not showing such regular progression in their outer
parts.

%%%%%%%%%%%%%%%%%%%%%
% Fig 6 trajectories
%%%%%%%%%%%%%%%%%%%%%
\begin{figure*}
   \centering
   \resizebox{\hsize}{!}{\includegraphics{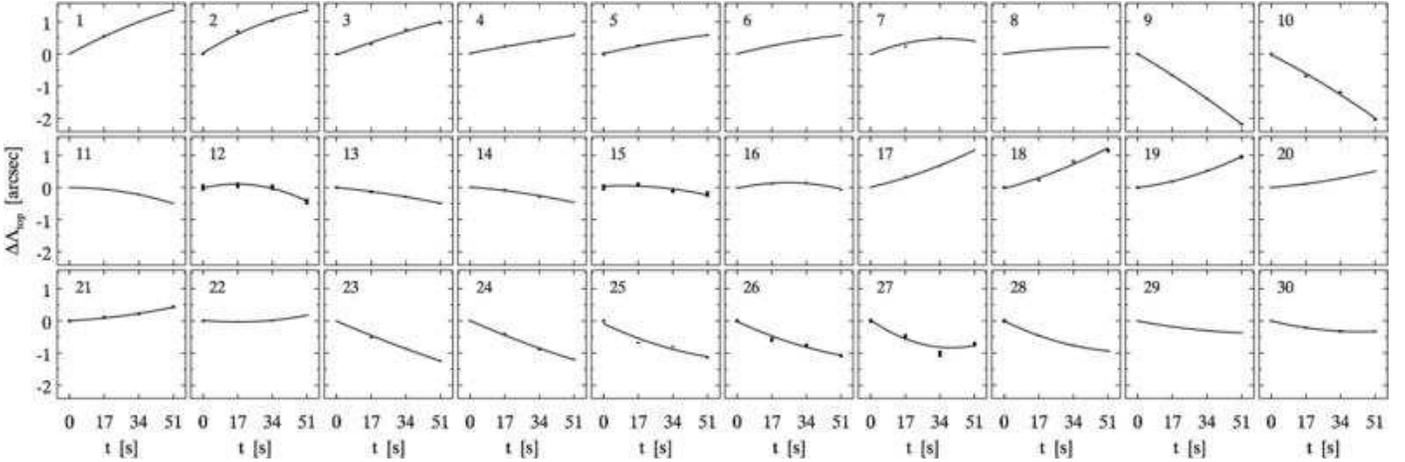}}
   \caption[]{
     Trajectory measurements $\Delta \Lambda_{\rm top}(t)$ of 30 jet
     tops along their axes. The heights of the symbols specify the
     error estimates. The solid curves are the parabolic fits. Nrs.\
     1--16: decelerating jets showing extension (1--8) and retraction
     (9--16). Nrs.\ 17--30: accelerating jets showing extension
     (17--22) and retraction (23--30).}
   \label{fig:trajectories}
\end{figure*}
%%%%%%%%%%%%%%%%%%%%%%%%%%%%%%%%%%%%%%%%%%%%%%%%%%%%%%%%%%%%%%%%%%%%%%%%

%%%%%%%%%%%%%%%%%%%%%%%%%%%%%%%%%%%%%%%%%%%%%%%%%%%%%%%%%%%%%%%%%%%%%%%
% Table - measurements
%%%%%%%%%%%%%%%%%%%%%%%%%%%%%%%%%%%%%%%%%%%%%%%%%%%%%%%%%%%%%%%%%%%%%%%%
\renewcommand{\arraystretch}{1.2}
\tabcolsep=7pt
\begin{table*}
\caption[]{
  \Lya\ DJ excursion amplitudes and kinematic quantities. The first
  number of each entry is the median value, followed by the mean with
  standard deviation (between parentheses) and the range (between
  angle brackets). The bottom row shows comparable determinations for
  \Ha\ DFs from \citet{DePontieuetal2007a}.
}
\label{tab:measurements}
\centering
\begin{tabular*}{\textwidth}{cccccc}
\hline \hline
%%%%%%%%%%%%%%%%%%%%%%%%%%%%%%%%%%%%%%%%%%%%%
%% header row
\multicolumn{2}{c}{feature type}  &
$A$  [arcsec]                     &
$|a|$  [m\,s$^{-2}$]              &
$v_{\rm max}$  [km\,s$^{-1}$]     &
$T$  [min]                        \\ % EOL

% 1st row
\hline
\multirow{2}{1.1cm}{\Lya\ DJs}                                            &
decelerating                                                              &
\multirow{2}{3.7cm}{$0.6 \;(0.8\!\pm\!0.5)\; \langle0.2,2.2\rangle$}      &  % A
$165 \;(235\!\pm\!173)\; \langle44,650\rangle$                            &  % dec for all decelerating
$13  \;(17\!\pm\!9)\; \langle6,35\rangle$                                 &  % vmax for decelerating
$2.8 \;(3.3\!\pm\!2.2)\;\langle0.9,7.5\rangle$                            \\ % T  EOL

% 2nd row
                                                                          &
accelerating                                                              &
                                                                          &
$246 \;(257\!\pm\!175)\; \langle87,797\rangle$                            &
$21 \;(18\!\pm\!7)\; \langle9,32\rangle$                                  &  % vmax for accelerating
not defined                                                               \\ % EOL

\hline

% 3rd row
\multicolumn{2}{c}{\Ha\ DFs}                                              &
      --                                                                  &
$136 \;(146\!\pm\!56)\; \langle40,320\rangle$                             &
$18 \;(18\!\pm\!6)\; \langle8,35\rangle$                                  &
$4.2 \;(4.8\!\pm\!1.4)\; \langle2,10.8\rangle$                            \\ % EOL
\hline
 \end{tabular*}
\end{table*}
%%%%%%%%%%%%%%%%%%%%%%%%%%%%%%%%%%%%%%%%%%%%%%%%%%%%%%%%%%%%%%%%%%%%%%%%%%%%%

The question then was how to define the jet top locations in order to
measure jet length variations in terms of their top displacements
$\Delta \Lambda_{\rm top}(t)$. For each DJ we selected intensity
thresholds at the onset of steep intensity decay with $\Lambda$ and at
the outer limit of visibility, as illustrated by horizontal dashes in
Fig.~\ref{fig:profiles}. Picking an intensity value from such a range
as the top location in one profile yields samples of $\Delta
\Lambda_{\rm top}(t)$ for the other three by finding the displacement
in $\Lambda$ for the outermost pixel with that intensity. We automated
this in a procedure using a random generator to pick threshold values
from a normal distribution centered on and covering the selected
range, and so obtained distributions of $\Delta \Lambda_{\rm top}(t)$
displacements for each jet that yield rms error estimates. This was
repeated using each of the four frames as initial reference; the four
error estimates were averaged. These errors represent the
threshold-definition uncertainty and the amount of gradient divergence
between the four profiles per jet. They are smallest when the curves
are parallel, as in the case of DJ~9 in the bottom-left panel of
Fig.~\ref{fig:profiles}.

Following the example of \citet{Hansteenetal2006} and
\citet{DePontieuetal2007a} we fitted the measured displacements, using
the error estimates as weights, with parabolas $\Delta \Lambda_{\rm
top}(t) = v_1\,(t-t_1) + (a/2)\,(t-t_1)^2$ where $v_1$ is the onset
velocity (positive when upward) at the time $t_1$ at which the first
frame was taken and $a$ the deceleration (when negative) or
acceleration (when positive). The terminal velocity $v_4$ at the time
of the fourth frame is $v_4 = v_1 + a\, (t_4-t_1)$ where $t_4-t_1 =
51$\,s.

Since the onset and terminal velocities $v_1$ and $v_4$ differ in all
cases and can be negative, we define a maximum velocity $v_{\rm max}$
as the higher of $|v_1|$ and $|v_4|$. The apex of parabolic
decelerated motion starting with $v_{\rm max}$ is reached after
$v_{\rm max}/|a|$ seconds. We estimate the total jet excursion
duration $T$ for decelerating jets by doubling this value: $T =
2v_{\rm max}/|a|$. Since the four frames are likely to sample only
part of a jet top trajectory, both $v_{\rm max}$ and $T$ represent
only lower limits.

Finally, we discarded 10 more DJs from our sample because the error
estimates for their $a$ and/or $v_{\rm max}$ determinations exceeded
100\%.

%%%%%%%%%%%%%%%%%%%%%%%%%%%%%%%%%%%%%%%%%%%%%%%%%%%%%%%%%%%%%%%%%%%%%%
\section{Results} \label{sec:results}
%%%%%%%%%%%%%%%%%%%%%%%%%%%%%%%%%%%%%%%%%%%%%%%%%%%%%%%%%%%%%%%%%%%%%%

Figure~\ref{fig:trajectories} shows the top trajectory measurements,
the corresponding error estimates, and the parabolic fits for the
remaining 30 \Lya\ DJs. Table~\ref{tab:measurements} specifies the
average values of the trajectory amplitudes and the fit parameters
together with their variances and ranges.

Most jet tops do not travel far during the short sequence duration,
only over a median distance of 0.6~arcsec, but these travel
measurements are nevertheless significant thanks to VAULT's 0.3-arcsec
resolution. Some of the decelerating jets have significantly greater
deceleration $-a$ than the solar surface gravity (274\,m\,s$^{-2}$);
we call these superballistic. The maximum velocity estimates $v_{\rm
max}$ all but one exceed the chromospheric sound speed
\citep[8~km\,s$^{-1}$, e.g.,][]{Uitenbroek2006}. The excursion
duration estimates $T$ range from 1 to 7.5 minutes with a median at
2.8~min. Again, we emphasize that the last two parameters are likely
to be underestimated when the brief image sequence did not cover the
full jet-top excursion.

\begin{figure}
   \resizebox{\hsize}{!}{\includegraphics{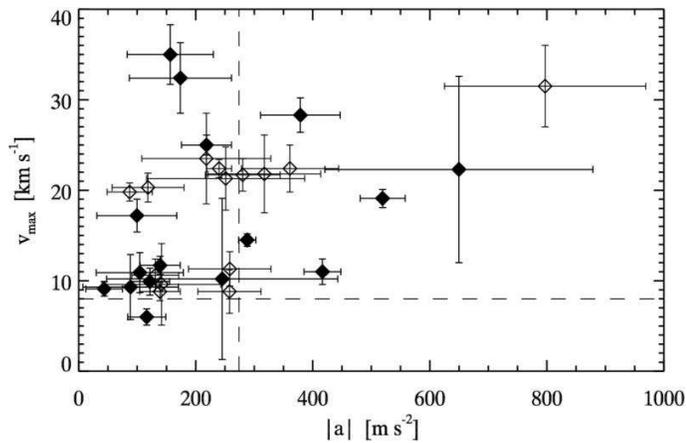}}
   \caption[]{
   Maximum velocity amplitude $v_{\rm max}$ against
   deceleration/acceleration amplitude $|a|$ for the selected 30 \Lya\
   DJs. {\em Filled symbols\/}: decelerating DJs. {\em Empty
   symbols\/}: accelerating DJs. {\em Dashed lines\/}: chromospheric
   sound speed of 8\,km\,s$^{-1}$ and solar surface gravity.
}  
   \label{fig:velocities}
   \end{figure}

Figure~\ref{fig:velocities} plots the maximum velocity amplitude
$v_{\rm max}$ against the deceleration/acceleration amplitude $|a|$
for the selected 30 \Lya\ DJs. The scatter and the errors are large
and the statistics low, but the plot suggests positive correlation
between $v_{\rm max}$ and $|a|$ for both the decelerating and the
accelerating \Lya\ DJs.

%%%%%%%%%%%%%%%%%%%%%%%%%%%%%%%%%%%%%%%%%%%%%%%%%%%%%%%%%%%%%%%%%%%%%%%%
% Fig 7 velocities
%%%%%%%%%%%%%%%%%%%%%%%%%%%%%%%%%%%%%%%%%%%%%%%%%%%%%%%%%%%%%%%%%%%%%%%%
%%%%%%%%%%%%%%%%%%%%%%%%%%%%%%%%%%%%%%%%%%%%%%%%%%%%%%%%%%%%%%%%%%%%%%%%

%%%%%%%%%%%%%%%%%%%%%%%%%%%%%%%%%%%%%%%%%%%%%%%%%%%%%%%%%%%%%%%%%%%%%%%%
\section{Discussion} \label{sec:discussion}
%%%%%%%%%%%%%%%%%%%%%%%%%%%%%%%%%%%%%%%%%%%%%%%%%%%%%%%%%%%%%%%%%%%%%%%%

Obviously, longer \Lya\ image sequences at the VAULT-II resolution are
needed to gain better information than these four images provide.
Nevertheless, they do show the existence of dynamic jets that fan out
from network and extend or retract during the one-minute image
sequence.
  
How do these \Lya\ DJs compare to the now well-studied and fairly
well-explained \Ha\ DFs? They have similar shapes, occupy similar
locations on the solar surface, show similar hedge-row morphology, and
extend or retract along their length the same way.
%RE thanks, the same way is better - they do so literally
The measurements for the 16 decelerating ones in
Table~\ref{tab:measurements} show fairly good correspondence with the
\Ha\ DF measurements listed at the bottom of the table.
Figure~\ref{fig:velocities} suggests a positive correlation between
$v_{\rm max}$ and $|a|$ comparable to the linear relations for \Ha\
DFs. Our estimate of the mean \Lya\ DJ excursion duration is about one
and half minutes shorter than the average \Ha\ DF periodicity reported
by \citet{DePontieuetal2007a}, but represents only a lower
limit. Three of the sixteen DJs show substantially greater
%RE this smaller > greater is beyond me. British? We write American...
superballistic deceleration than the maximum of 320\,m\,s$^{-2}$ for
DFs reported by \citet{DePontieuetal2007a} and the maximum of
400\,m\,s$^{-2}$ reported by \citet{RouppevanderVoortetal2007} and
\citet{DePontieuetal2007c} for quiet-Sun mottles and for type-1
spicules, respectively, but with large uncertainty for one of them. We
conclude that, overall, our decelerating \Lya\ DJs are rather similar
to \Ha\ DFs.

However, in contrast to the \Ha\ DF studies, we found a roughly equal
number of \Lya\ DJs that significantly accelerate rather than
decelerate. These are collected in the lower half of
Fig.~\ref{fig:trajectories}. The two sets of trajectories together
suggest that they might actually be incomplete samples of sinusoidal
motion or, as the referee has pointed out, sample the end of one
shock-driven feature and the
%RE didn't understand the change
onset of the next one if these \Lya\ DJs occur in succession as \Ha\
DFs do. Note that our length estimates do not differ significantly
between the two sets, but these are rather arbitrary.
%RJ I meant the same with my sinusoidal but it is true that the DFs have
%RJ sharp v minima. But in both you would expect the accelerating ones to be
%RJ shorter on average. 

An obvious difference between \Lya\ DJs and \Ha\ DFs is that the
former appear bright against dark backgrounds, the latter dark against
bright backgrounds. \Lya\ emissivity indicates the presence of gas
that is sufficiently hot for collisional excitation of its upper level
at 10.2~eV. Sufficient population of that level is also required to
make fibrils opaque in \Ha. The recent study by
\citet{Leenaartsetal2007a} suggests that cool fibrils may maintain
much greater \Ha\ opacity than their temperature would suggest if they
have recently undergone shock heating. Hence, \Ha\ DFs may remain
opaque also when they are cool during part of their lifetime. They
will be much more opaque in \Lya. Can the same fibril appear bright in
\Lya\ and dark in \Ha?  Perhaps through hot-sheath topology of the
transition region, with the latter surrounding cool DFs as hot shells
\citep[e.g.,][]{Rutten2007,DePontieuetal2007a}. The neutral-hydrogen
diffusion mechanism recently proposed by \citet{Judge2008} seems a
viable candidate. Detailed time-dependent numerical MHD simulation may
demonstrate this. However, realistic evaluation of \Ha\ formation from
such simulations remains a formidable challenge that may be helped by
constraints derived from \Lya.

Finally, the \Lya\ DJs studied here are likely related to
chromospheric phenomena such as regular spicules
\citep[e.g.][]{Sterling2000}, straws \citep{Rutten2006,Rutten2007},
and type-2 spicules \citep{DePontieuetal2007c,Langangenetal2008c}.

\begin{acknowledgements}
  We are indebted to P.~S\"utterlin for providing his image-alignment
  routines and to the referee whose comments improved the paper
  substantially. J.\,Koza thanks J.\,Ryb\'ak, P.\,G\"om\"ory, and
  A.\,Ku\v{c}era for valuable discussions and comments. His research
  was supported by EC Marie Curie European Re-Integration Grant
%RE no the
  MERG-CT-2007-046475 and by the Slovak Research and Development
  Agency APVV under contract APVV-0066-06. The VAULT development was
  supported by the Office of Naval Research (task area SP033-02-43)
  and NASA (defense procurement request S-84002F).
\end{acknowledgements}

\end{document}